\documentclass[doublecolumn]{rmaa}
\usepackage{epsfig}

\newcommand{\oi}{[O~{\sc i}]6300~\AA}

\def \ha   {H$\alpha$}
\def \oiii {[O~{\sc iii}]5007~\AA}
\def \nii  {[N~{\sc ii}]6584~\AA}
\def \NII  {[N~{\sc ii}]6548~\&~6584~\AA}
\def \sii {[S~{\sc ii}]6717 \& 6731~\AA}
\def \heii  {He{\sc ii}~6560~\AA}
\def \heiis  {He{\sc ii}~4686~\AA}
\def \tena#1 #2 {\ifmmode{#1 \times 10^{#2}}\else{$#1 \times 10^{#2}$}\fi}
\def \kms  {\ifmmode{~{\rm km\,s}^{-1}}\else{~km s$^{-1}$}\fi}
\def \vhel {\ifmmode{~V_{{\rm HEL}}}\else{~$V_{{\rm HEL}}$}\fi}
\def \vsys {\ifmmode{~V_{{\rm SYS}}}\else{~$V_{{\rm SYS}}$}\fi}

\def \deg  {\ifmmode{^{\circ}}\else{$^{\circ}$}\fi} 
\def \msun {\ifmmode{{\rm\ M}_\odot}\else{${\rm\ M}_\odot$}\fi}
\def \myr  {\ifmmode{{\rm\ M}_\odot{\rm\ yr}^{-1}}
         \else{${\rm\ M}_\odot$ yr$^{-1}$}\fi}
\def \etal {et al.\ }

   \title{The global kinematics of the Dumbbell planetary nebula
   (NGC 6853, M27, PN G060.8-03.6)}

   \author{ J. Meaburn\affil{ Instituto de Astronomia, UNAM,
   campus Ensenada}
   \and  P. Boumis\affil{Institute of Astronomy \& Astrophysics, 
   National Observatory of Athens, Greece.}
   \and  P. E. Christopoulou, C.~D.~Goudis\affil{Astronomical Laboratory, 
   University of Patras, Greece}
   \and  M. Bryce\affil{ Jodrell Bank Observatory, UK}
   \and J. A. L{\' o}pez\affil{Instituto de Astronomia, UNAM, campus Ensenada}}

\fulladdresses{
\item First Author :  Instituto de Astronomia, UNAM, Apdo. Postal 877. 
Ensenada, B.C.
\item Second Author : Institute of Astronomy \& Astrophysics, National 
Observatory of Athens, 
I. Metaxa \& V. Paulou, GR--152 36 P. Penteli, Athens, Greece.}

\shortauthor{J. Meaburn}
\shorttitle{Dumbbell Nebula}

\resumen{Se han obtenido perfiles resueltos espacialmente 
de las l\'{\i}neas de
emisi\'on de \heii, \ha\ y \NII\ sobre dos
largos cortes ortogonales en la nebulosa planetaria conocida como
Dumbbell. Se muestra que el vol\'umen central, brillante, que
emite \heii\ es particularmente inerte, con velocidades de
expansi\'on de $\leq$ 7 \kms. Esta zona est\'a envuelta por un
cascar\'on que emite en \oiii\ y que se expande a 13 \kms\
y otro cascar\'on externo, tambi\'en de \oiii,
expandi\'endose a 31 \kms. Un siguiente cascar\'on externo que
emite en \nii, se expande a 35 \kms. Se presenta una nueva
imagen de \ha\ + \NII, con cont\'{\i}nuo substraido, de
campo amplio que abarca el halo de 15' de di\'ametro y se compara con
recientes medidas de movimientos propios de la estrella central y los
perfiles de l\'{\i}neas de emisi\'on de la secci\'on interior del halo,
presentados aqu\'{\i}. Estos datos suguieren interacci\'on de la nebulosa
con el medio ambiente interestelar. El cascar\'on brillante que emite en
\nii\ debe estar penetrando el halo, que es relativamente inerte,
con una velocidad diferencial  $\geq$ 25 \kms. Los resultados
aqu\'{\i} presentados se comparan con modelos actuales
para la creaci\'on de nebulosas planetarias.
}

\abstract{Spatially resolved profiles of the \heii, \ha\ and \NII\
nebular emission lines have been obtained in two orthogonal
long cuts over the Dumbbell planetary nebula. 

The central \heii\ emitting  volume of the bright 
dumbbell structure 
is shown to be
particularly inert with an expansion velocity of $\leq$~7~\kms.
This is enveloped by an inner \oiii\ emitting shell expanding at 13~\kms,
an outer \oiii\ shell expanding at 31~\kms\ which is on the inside of
the outer \nii\ emitting shell expanding at 35~\kms. 
A new \ha\ + \NII\ continuum--subtracted image of the 15\arcmin\ diameter
halo has also been compared with recent proper motion measurements
of the central star and the present line profiles from the halo's
inner edge. Interaction with the ambient interstellar medium
is suggested. The bright \nii\ emitting shell must be running
into this relatively inert halo with a differential velocity
of $\geq$~25~\kms.

The present results are compared with currents models for the creation
of planetary nebulae.}

\keywords{planetary nebulae:individual:Dumbbell, 
NGC~6853 -- ISM: kinematics and
dynamics}

\begin{document}
\maketitle

%

\section{Introduction}

NGC 6853 (M27, PN G060.8-03.6) is a planetary nebula (PN) of large
angular size, more commonly known as the Dumbbell Nebula. Its distance
has been measured by parallax observations as 360 $\pm$ 60 pc 
 (Pier et al 1993, Harris et al 1997)
and 420 $\pm$ 60 pc (Benedict et al 2003),
making it one of the closest PNe.

This object has been classified morphologically by Balick
(1987) as late elliptical, by Chu et al (1987) as a Type
I multiple shell planetary nebula, and by Manchado et al (1996)
as elliptical with inner structure and multiple shells.

The elliptical main body of the nebula has an angular size of
8\arcmin~$\times$~5\arcmin\ along the major and minor axes at position
angles (PA)~125\deg\ and ~35\deg\ respectively, with the maxima of
brightness at the ends of the minor axis (the bar--like feature which
crosses the nebula from NE to SW).

The main nebula was reported to contain a so--called second internal
shell (Maury \& Acker 1990) of angular dimensions 1.\arcmin 2
$\times$ 0.\arcmin 8, visible in the UV (Maury \& Acker 1990) and
in the \sii\ emission lines (Moreno--Corral et al 1992).
Also it is surrounded by an extended (15 arcmin across)
fainter halo first
reported by Millikan (1974). Balick et al (1992),
Moreno--Corral et al (1992) and Papamastorakis et al
(1993) have all presented narrow band images of the Dumbbell showing this
extensive halo and noted the existence of radial `rays'
connecting the halo and the main nebula. The recent deep images of
Manchado et al (1996) show all these features very
clearly.

The central star of Dumbbell nebula is classified as O7 (Cerruti--Sola 
\& Perinotto
1985) and its
estimated temperature and luminosity are characteristic of an evolved
planetary nebula nucleus. Absolute parallax
and relative proper motion for this star have recently been presented
by Benedict et al (2003). The central star is also responsible
for the reported X--ray emission of the nebula (Chu et al 1993)
for its surface temperature is $\approx$ 10$^{5}$ K (Napiwotzki
1999).

Zuckerman \& Gatley (1988) mapped the molecular hydrogen
emission and found that it followed the brightest regions of optical
emission. Huggins et al (1996) have detected CO emission and
find a complex filamentary structure to the molecular gas. Spectral
observations at a point 68\arcsec\ S, 63\arcsec\ W of the central star
(coinciding with the brightest optical emission) show line splitting
of 30\kms. Kastner et al (1996) present a detailed H$_{2}$ map
again showing the complex structure of the neutral gas. Dusty globules
of molecular gas in the near side of the outer \nii\ and
\oi\ emitting shell of the central bar of the nebula were discovered
by Meaburn \& L{\'o}pez (1993), silhouetted against the more
central \oiii\ and \heiis\ emission, suggesting that they have been
ejected in the post--asymptotic phase of the central star.

Goudis et al (1978) obtained insect--eye Fabry--Perot
interferograms of the \oi, \oiii\ and \nii\ emission lines from the
Dumbbell. They found that the lower excitation \oi\ and \nii\ emission
could be explained by a cylindrical structure, expanding radially with
respect to its axis with an expansion velocity of $\sim$30\kms,
although they note that the \oiii\ emission from near the exciting star
indicates more complex motions. More recent observations of the \oiii\
emission (Meaburn et al 1992) obtained from the very central
region of the Dumbbell revealed the presence of four distinct velocity
components within each emission line profile, indicating the existence
of an outer shell expanding at 31\kms\ (probably connected to the
low--ionisation structure) and the suggestion of 
an inner shell expanding radially at
12\kms (Meaburn et al 1992). Such a double--shell 
structure in the light of \oiii\ has also
been observed in the Helix nebula (Meaburn et al 1998, O'Dell et
al 2004).

In the present  paper, longslit spectroscopic observations of the Dumbbell
nebula obtained with high spectral and spatial resolution, across both
the main nebular shell and fainter  halo, are presented. 
A new deep wide-field image, with continuum subtracted, permits
comparison with both the kinematical features of the halo that are 
presented here and a 
recent proper motion measurement of the central star by Benedict (2003).
The spatio--kinematical structure of the various
components of the Dumbbell nebula can be compared in a significant way 
with the
predictions of dynamical models.

\section{Observations and results}

\subsection{Deep Imagery}

The wide--field image of Dumbbell nebula shown in Figs. 1 a \& b, 
was obtained with the
0.3 m Schmidt--Cassegrain telescope at Skinakas Observatory, Crete,
Greece on June 13, 2004. The observations were performed with a 1024
$\times$ 1024 Thomson CCD
which provides a 70\arcmin\ $\times$ 70\arcmin\ field of view and an
image scale of 4\arcsec\ per pixel. The Dumbbell nebula 
was observed for 4800 s
in total through the \ha\ plus \NII\ filter and for 180 s with 
the corresponding
continuum filter. The latter continuum image was subtracted from the 
former line emission image 
to eliminate the
confusing star field (more details of this technique can be found in
Boumis et al 2002). The astrometric solution was made using
the HST
Guide star catalogue (Lasker et al 1999). The equatorial
coordinates quoted in this work refer to epoch 2000.
Standard IRAF and MIDAS routines were employed for the reduction of the
data.
All frames were bias subtracted and flat-field corrected using a series
of
well exposed twilight flat--fields.

\begin{figure*}
\epsfclipon
\centering
\mbox{\epsfysize=7in\epsfbox[0 0 418 769]{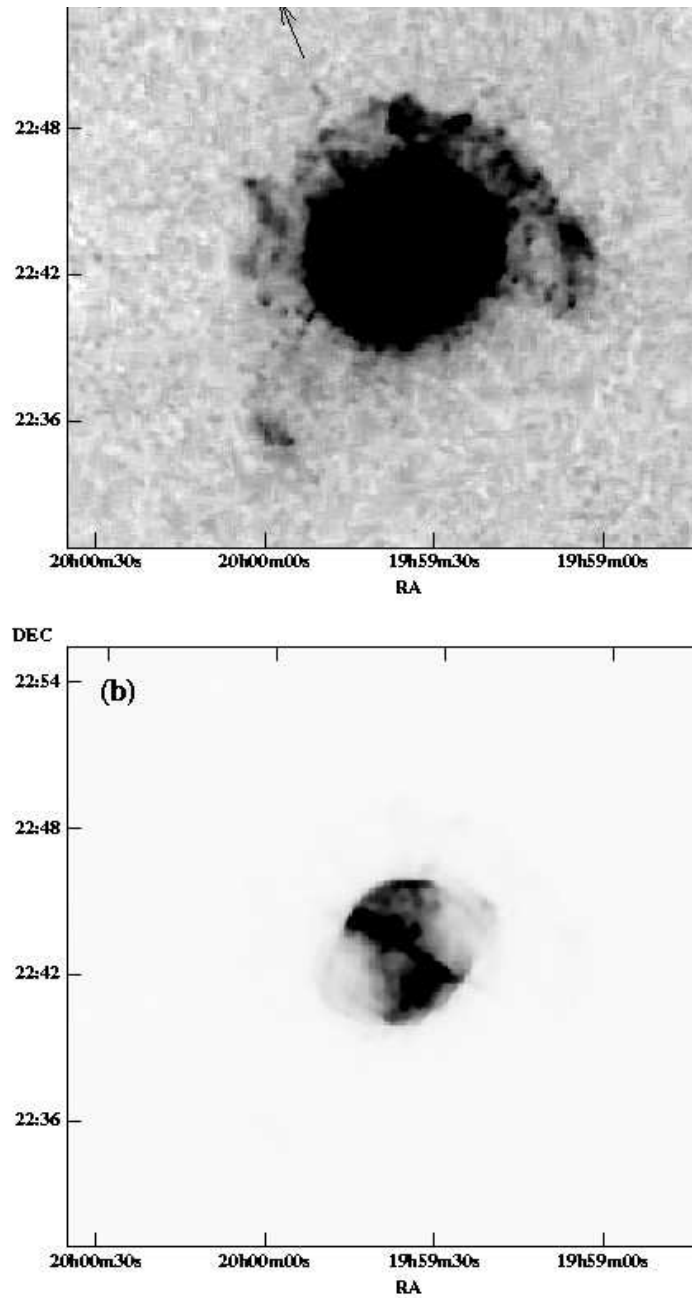}}
\caption{a) A deep representation of the continuum subtracted
\ha\ plus \NII\ image. This is an enlargement 
from the whole 70\arcmin\ x 70\arcmin\ field.The arrowed line
indicates the proper motion found for the central star. b) 
A lighter representation of the same image
shown in a.}
\label{fig01}
\end{figure*}

\subsection{Spectroscopy}
Spatially resolved profiles of the \heii\, \nii\ and \ha\ nebular
emission lines along five slit positions were obtained
on 19--21 Nov. 1987 (as yet unpublished)
 at high
spectral resolution with the Manchester Echelle Spectrometer (MES;
Meaburn et al 1984) combined with the f/8 (converted) focus of
the 4.2~m William Herschel Telescope (WHT).

MES was used in its primary mode with a narrow band 100 \AA\ bandwidth
interference filter isolating the 87$^{\rm th}$ echelle order. The
IPCS--CCD detector was used with 260 increments (each of 0.\arcsec75) along
the slit length (195\arcsec) and 1260 data receiving channels in the
dispersion direction. The slit width was 150 $\mu$m to give a spectral
resolution of 11 \kms.

Three of the slit positions were orientated along the SW--NE direction
of the nebula (effective angular length on the sky ~8\arcmin.4) as
marked in Fig. 2 whereas the other three were employed in
the NW--SE direction (angular length on the sky ~8\arcmin.4) are
also marked in Fig. 2.

\begin{figure*}
\resizebox{\hsize}{!}{\includegraphics{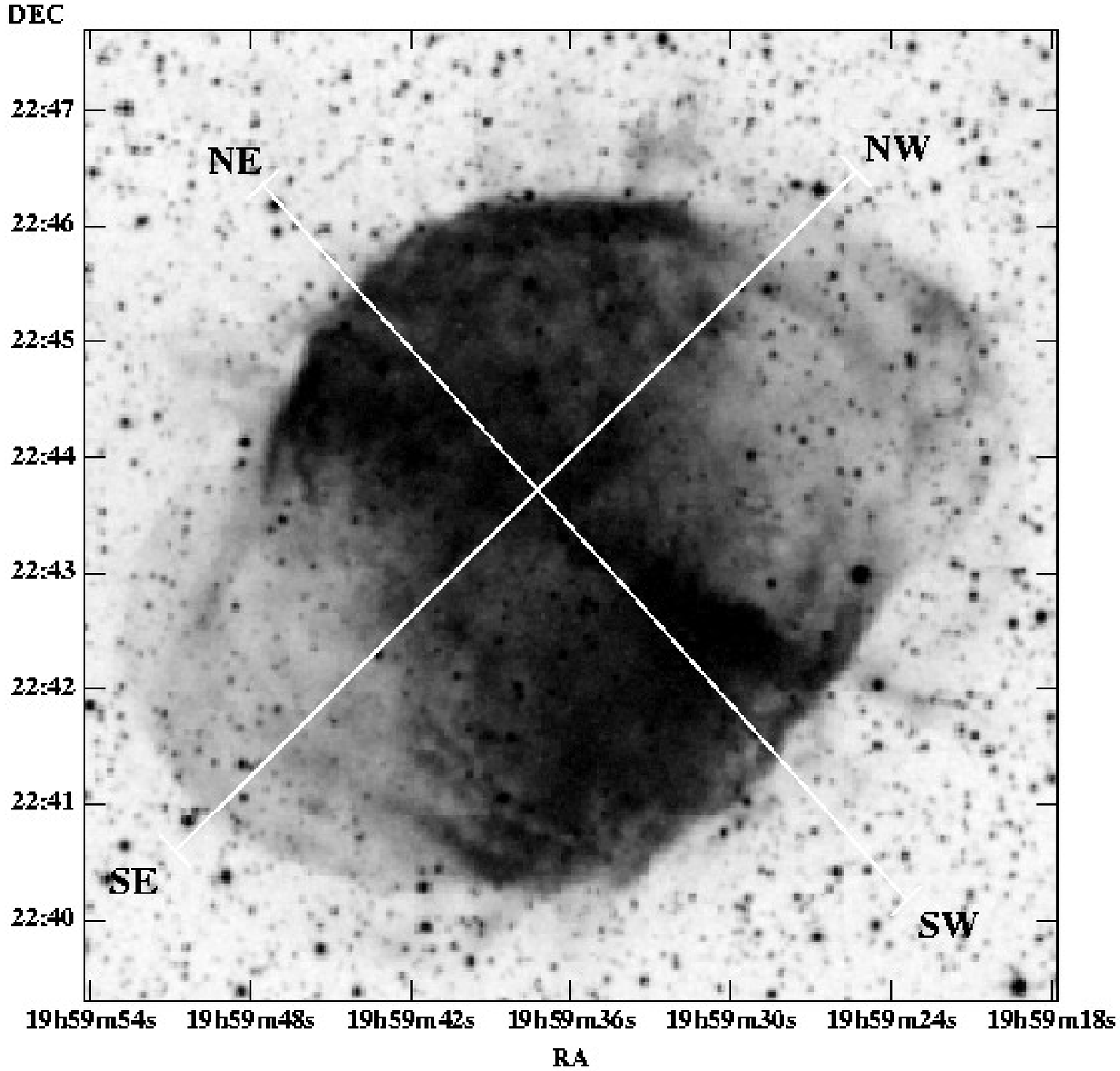}}
\caption{The NW--SE  and SW--NE slit positions where profiles
were obtained,
are shown against a 8.\arcmin5 $\times$ 8.\arcmin5 DSS2 Red image of the
Dumbbell nebula.}
\label{fig02}
\end{figure*}

The resultant spectra were calibrated to $\pm$~1\kms\ accuracy against
the reference spectra of a CuAr lamp. The reduction of the data was
performed on the Manchester node of the UK STARLINK computer network
using the FIGARO, KAPPA, CCDPACK packages. Each sequence of the
separate data sets was combined into a single mosaic in each direction
using the CCDPACK routine MAKEMOS in order to eliminate possible
discontinuities caused by the different effective exposure times and
background levels of the slits.

Greyscale representations of the final mosaics of the position -- velocity
(pv) arrays of profiles along the SW--NE and
NW--SE directions of the nebula are presented in Fig. 3 for both \ha\
and \nii\ emission lines.

\begin{figure*}
\epsfclipon
\centering
\mbox{\epsfysize=7in\epsfbox[20 20 575 753]{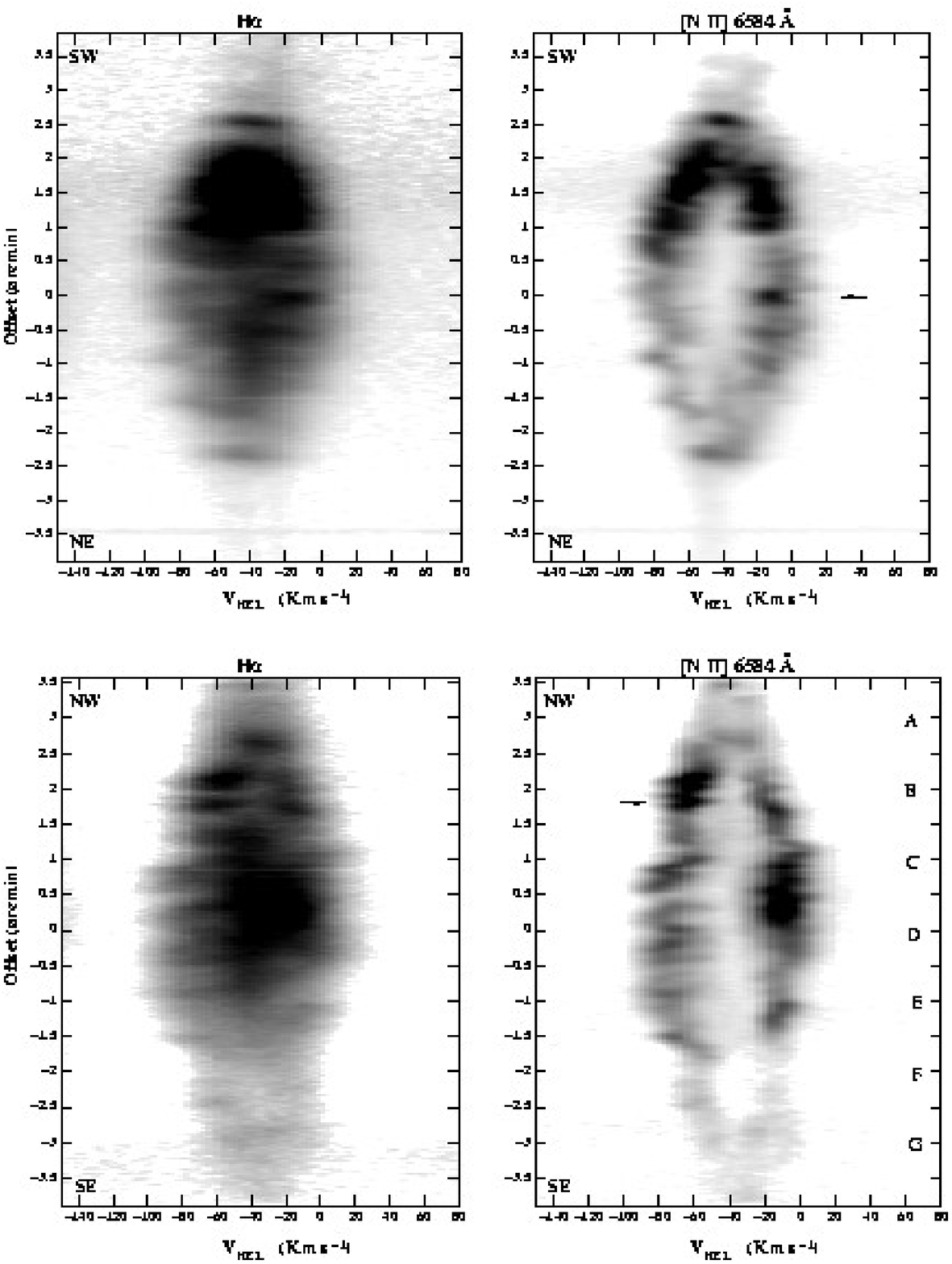}}
\caption{Position velocity (pv) arrays of the \ha\ and \nii\
profiles from slits 1 and 2 are shown in the top and bottom
images, respectively. The horizontal axes are corrected to heliocentric
radial velocity (\vhel) and the vertical axes are in arcmin and
correspond spatially to the solid lines  in Fig. 1.}
Two examples of the `V--shaped' velocity features are arrowed.
The \nii\ line profiles shown in Fig. 5 were extracted
from the positions A--G (each cut 3 arcsec wide) marked against
the \nii\ greyscale for the NW--SE slit (see Fig.~2).
\label{fig03}
\end{figure*}

\begin{figure*}
\epsfclipon
\centering
\mbox{\epsfysize=7in\epsfbox[0 0 591 841]{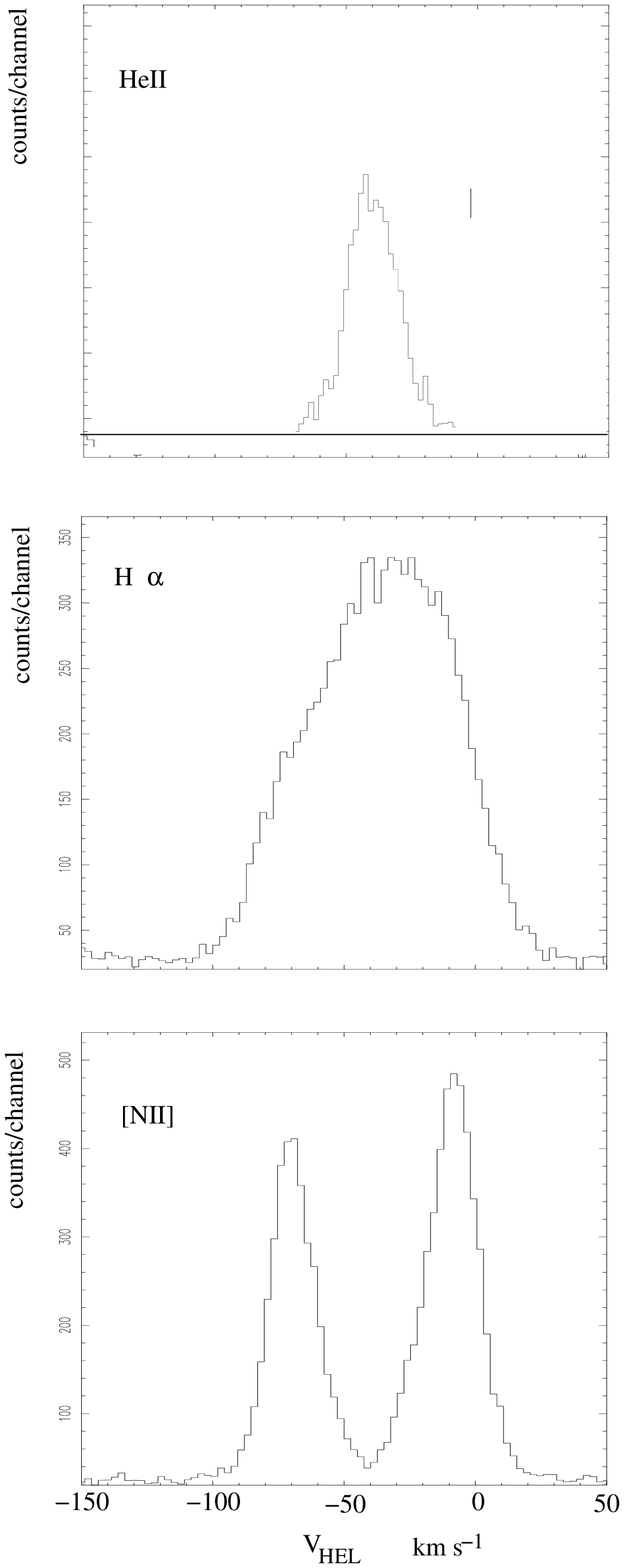}}
\caption{\ha\ and \nii\ line profiles from a 1 arcsec long increment
of the slit length over the nebular core i.e. at 0\arcmin\ position
for the NW--SE slit in Fig. 3 are compared with the profile of the 
\heii\ emission line but extracted, because of its faintness, 
for a 20\arcsec\ width
centred on the same position. The counts/channel for this \heii\
profile are unreliable and therefore omitted but its peak surface
brightness is around 0.05 that of the \ha\ line.}
\label{fig04}
\end{figure*}

\begin{figure*}
\epsfclipon
\centering
\mbox{\epsfysize=7in\epsfbox[0 0 594 765]{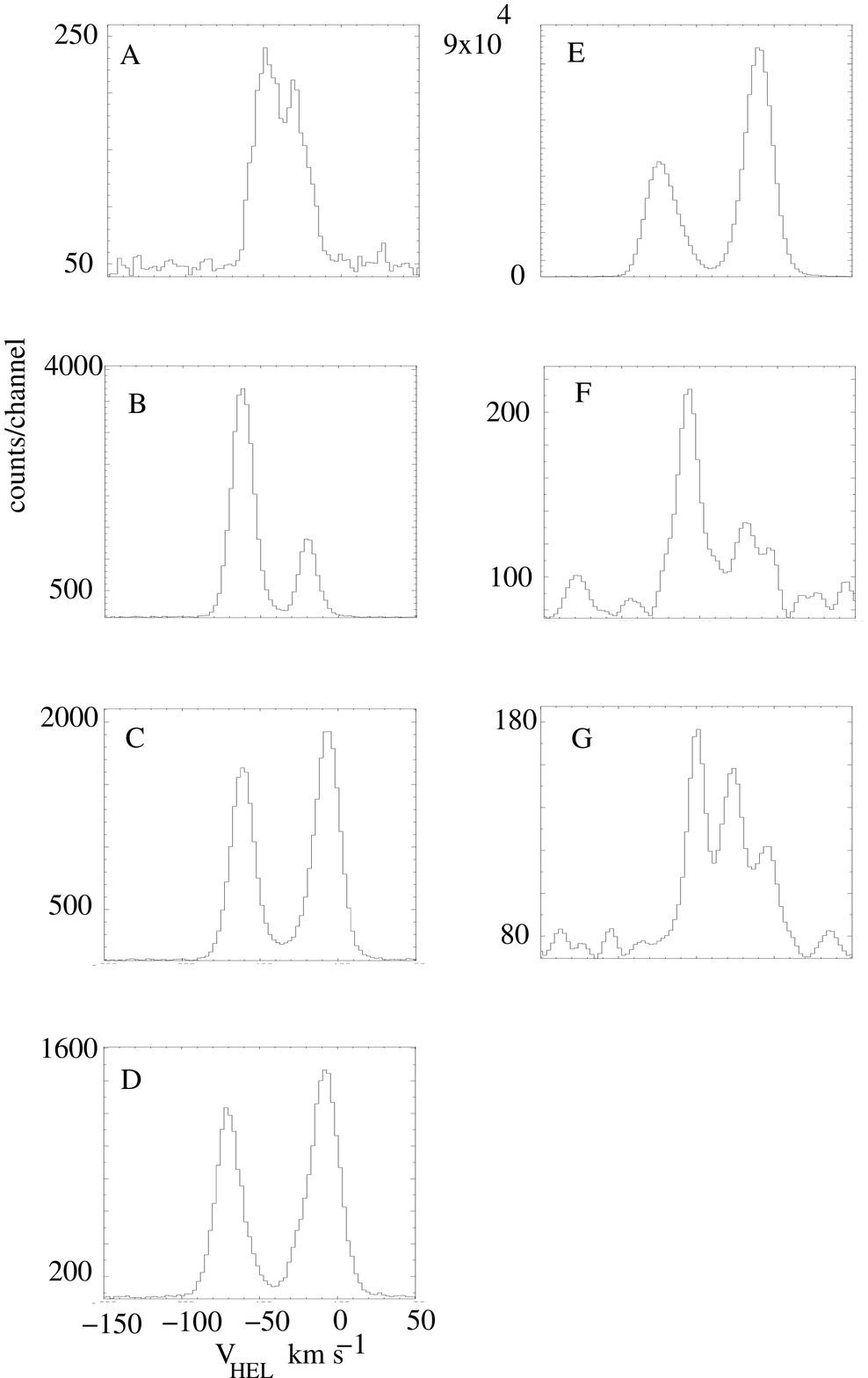}}
\caption{Sample \nii\ line profiles from  3 arcsec long increments
along Slit 1 from the position marked A--G in Fig. 3}
\label{fig05}
\end{figure*}

\begin{figure*}
\resizebox{\hsize}{!}{\includegraphics{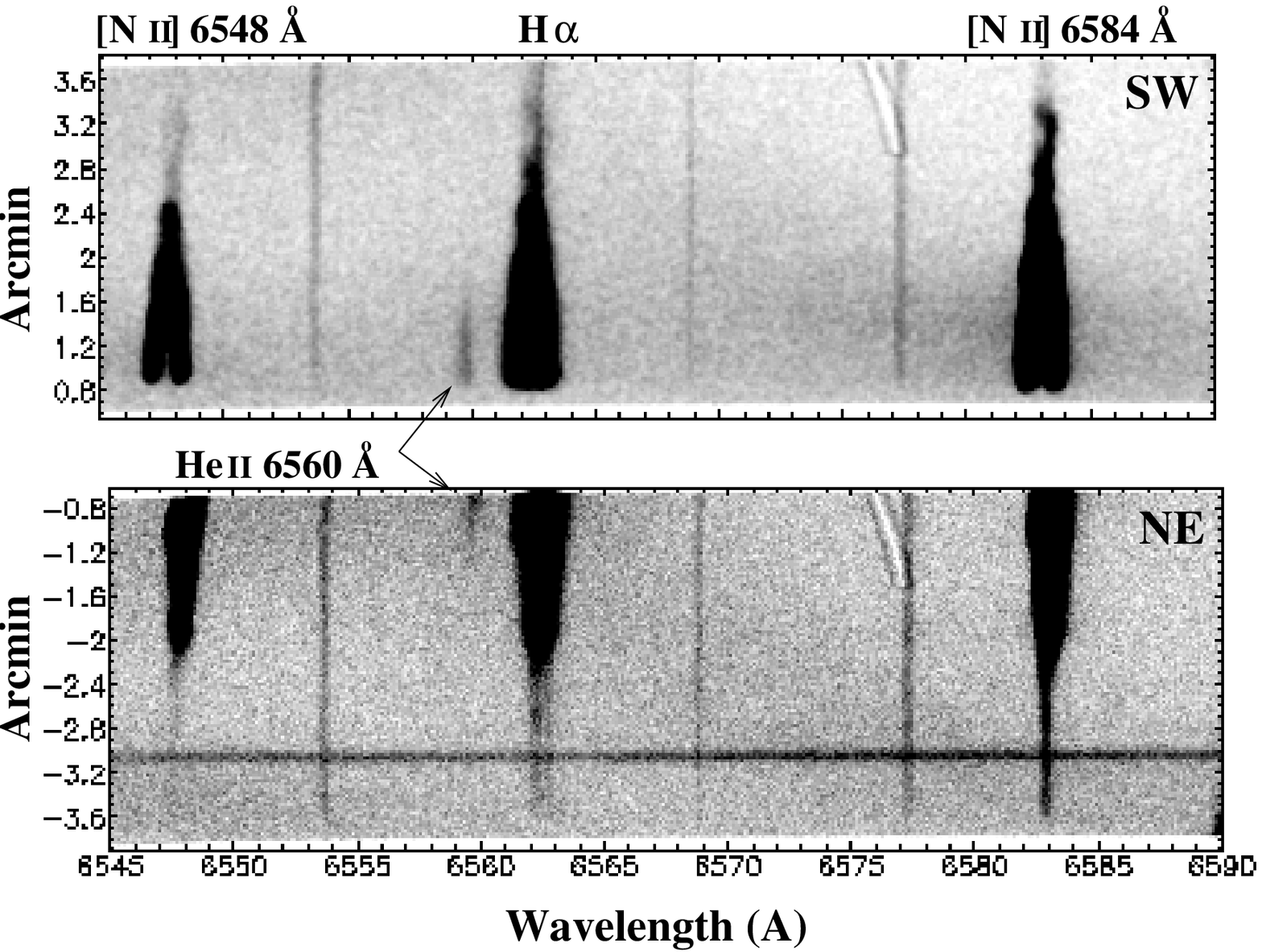}}
\caption{Negative grey--scale representations of the pv arrays of \heii, 
\ha\
and \NII\ profiles are shown for only the end slit lengths of
the SW--NE line of measurments in Fig. 2.   
The \ha\ and \nii\ line profiles from the faint PN halo can be
seen at the top end of the SW slit and the bottom end of the NE one. 
The dark horizontal band in the second panel is the continuous
spectrum of a star image which has also fallen on the slit. 
An instrumental artifact and airglow spectra  are also present}
\label{fig06}
\end{figure*}

\begin{figure*}
\resizebox{\hsize}{!}{\includegraphics{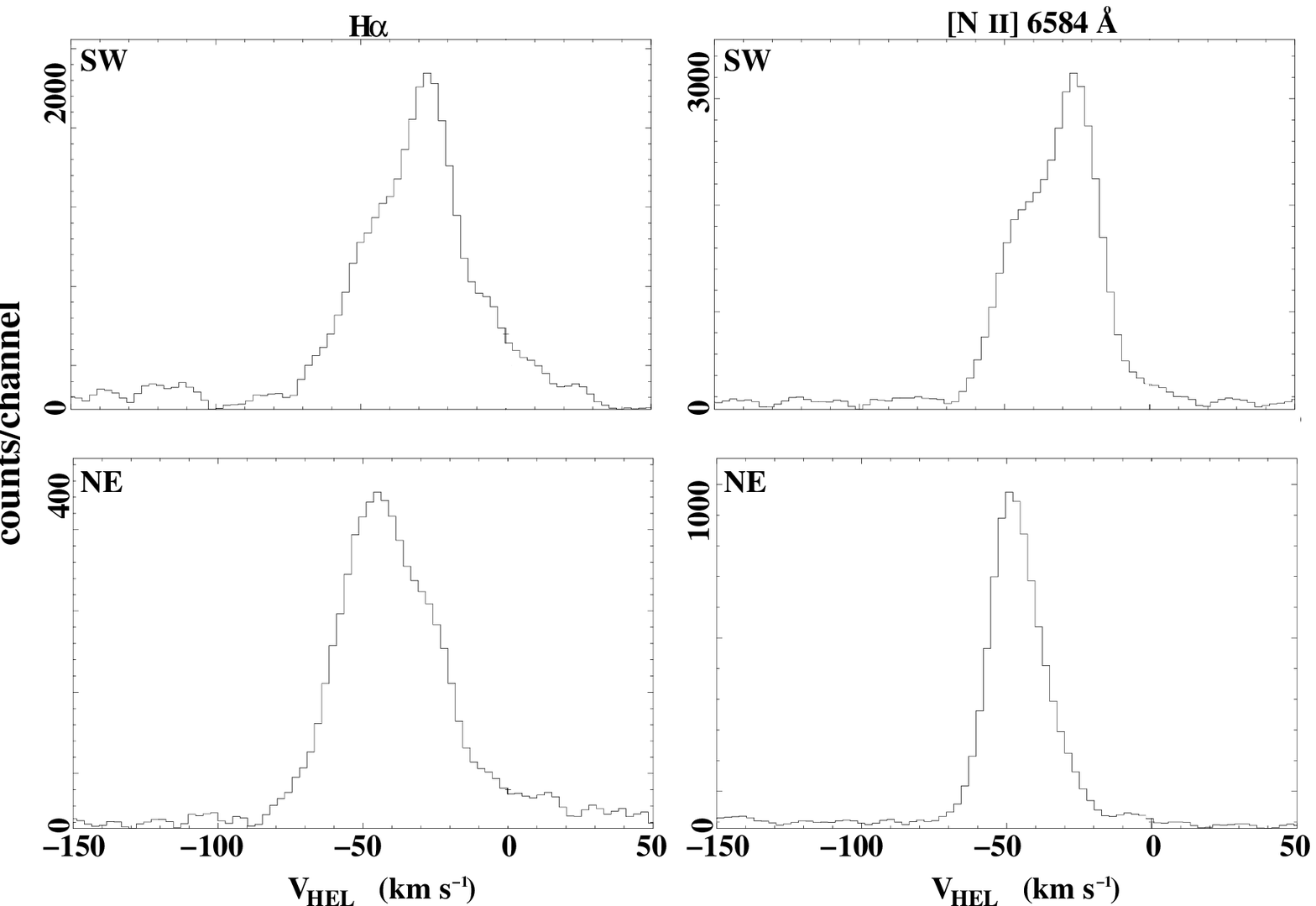}} 
\caption{\ha\ and \nii\ line profiles of the halo from the 
pv arrays in Figs. 3 \& 6, are presented. The halo profiles  
have been obtained for two lengths of the slit. Those designated
SW are for the top
36\arcsec\ length of the slit and those designated NE are for the
bottom 48\arcsec length.}
\label{fig07}
\end{figure*}

Profiles from the central increment (marked 0 arcmin in the
NW--SE arrays in Fig. 3) of the \heii, \ha\ and \nii\ emission
lines are compared in Fig. 4. \nii\ line profiles from the positions marked 
A--G along the NW--SE array (see Fig. 3) are shown in Fig. 5. 
These are for 3\arcsec\
wide lengths along the array each centred on these positions.

Deep, negative, greyscale representations of the parts of the pv arrays
(\heii, \ha\ and \NII) that include the faint halo of NGC~6853 are shown in
Fig.~6. \ha\ and \nii\ line profiles for the halo emission are shown
in Fig. 7.  
 
\section{Discussion}

\subsection{New kinematical features}

The most striking feature of the pv arrays of profiles from the main
body of NGC~6853 in Figs. 3 \& 4 is that the \ha\ emission appears to
originate within a large fraction of the nebular volume 
whereas \nii\ emission comes from an
outer shell expanding at 35~\kms\ with a systemic heliocentric  radial
velocity of \vsys\ =~ $-$41~\kms. Even though the thermal width
of the \ha\ line at 10,000 K ($\equiv$~21.4 \kms\ which
dominates the intrinsic broadening by the fine structural
components of the \ha\ line -- see below) is 3.7 times greater
than that for \nii\ it alone cannot explain the large differences between
the \nii\ and \ha\ profiles in Fig.~4. After all the \nii\ velocity
components are separated by 70~\kms.  
Moreover, an unpublished image of the the \heiis\ emission (O'Dell - 
private comm.) 
of  NGC~6853 shows that it originates in a diffuse central 
volume, 180\arcsec\ x 160\arcsec\ in size where the long axis
is aligned with the bright central bar of the nebula. This image was taken
with the same system employed for NGC~7293 (O'Dell 1998).  
The observed 
width of the central \heii\ 
profile in Fig.~4 has been measured in the present work
as 29.5~$\pm$~
1.3~\kms by fitting a single Gaussian.

The \heii\ line is composed of  nineteen fine structural components covering
the 6559.769--6560.209 \AA\ range. 
The intrinsic width of the \heii\
profile as a consequence can be approximated by a single Gaussian of 
$\geq$ 10~\kms\ width (the brightness distribution
will not be the same for all components; e.g. the seven components
that constitute the \ha\ line lie between 6562.709--6562.909 \AA\
but can be reasonably simulated by a single Gaussian of width
0.14~\AA\ i.e. $\equiv$ 6.4~\kms). 
 When the intrinsic width of the \heii\ line is 
combined with an 11~\kms\ wide instrumental profile and a 10.7~\kms\
thermal width (at 10,000 K) this leaves the turbulent motions
combined with any global expansion of the \heii\ volume to
contribute only $\leq$~12\kms\ to the observed width
of the central line profile. If turbulent motions within this 
central volume are around the sound speed (say 10 \kms) 
then any  radial expansion of the central \heii\ emitting volume
must then be $\leq$ 7~\kms. Note that the inner \oiii\ emitting shell,
whose existence is inferred by spatially limited observations,
has a measured expansion velocity of only 13~\kms\ (Meaburn et al, 1992).

In summary, the complex \ha\ profile in Fig.~4 must originate in the outer
\nii\ emitting shell expanding at 35 \kms, the \oiii\ emitting shell
on its inside surface expanding at 31 \kms\ (Meaburn et al 1992),
the inferred inner \oiii\ emitting shell expanding at 13 \kms\ (Meaburn et al
1992) as well as in the inner central \heii\ emitting  volume 
whose expansion, if 
any exists, is shown here to be $\leq$~7 \kms.The resultant broad
and complex \ha\ profile in Fig. 4 is consequently  the  
emission in different
proportion from all of these regions along this sight line 
convolved with the larger
thermal width of the \ha\ line.

Details of interest in
these pv arrays are the `V-shaped' features (two examples are arrowed
in Fig. 3) about 20\arcsec\ across and extending in radial velocity
beyond that of the expanding shell.

The \nii\ profiles along the SW--NE pv array describe a reasonable
`velocity ellipse' indicating circular radial expansion, whereas those
along the NW--SE length are nearly open--ended. A barrel--shaped or
quasi ellipsoidal,
expansion is indicated.

The \nii\  profiles from the SW end of the slit
in Figs. 6 \& 7  cover the inner
part of the faint 
outer halo and exhibit (see Fig. 1) 
two velocity components separated by about 25~\kms\ centred on the nebular
\vsys. Perhaps more significantly those from the NE end 
are narrow (20 \kms) and single.
Any expansion of the halo must then be $\leq$~13~\kms.
Within this interpretation the inner bright \nii\ emitting shell must
be running into this halo with a differential velocity of $\geq$~22
\kms.

It is interesting that Benedict et al (2003) give the proper motion
of the central star as 19.2 $\pm$ 1 mas yr$^{-1}$ along 
PA 21 $\pm$ 1\deg\ (see arrowed 
line in Fig. 1a) which at a
distance of 420 pc. indicates a tangential velocity of 37 \kms. 

\subsection{Dynamics}

Planetary nebulae (PNe) are formed by mass loss from stars with initial
masses $<$8 M$_{\odot}$. Since fast stellar winds with terminal
velocities of 600--3500\kms\ have been detected in a large number of PN
nuclei and since the progenitors of PNe have lost mass via slow winds at
the red giant phase and via `superwinds' on the asymptotic giant branch
(AGB) at 10--25\kms, the fast stellar winds inevitably will catch up
and interact with the previous slow winds.
The standard interacting winds model (Kwok, Purton \& Fitzgerald, 1978; Khan 
and West, 1985, Chu
et al 1993, Frank (1999), Balick \& Frank 2002) has provided an elegant 
explanation for the formation of the nebular structure in planetary 
nebulae during the transition of the stellar core from the post-AGB to 
the white dwarf stages. 

In these models the PN system consists of a central star, an energetic
stellar wind, an expanding ionized shell and the remnant envelope of
the progenitor red giant. The observations show that most PNe can
basically be described in terms of the above model but there are some
difficulties for complex PN with multiple shell components where
peculiar ejection processes, related to the evolution and/or the
nature of the central star, have been invoked (Miranda \& Solf
1992).

The pv arrays shown in Figs 3 and 6 immediately reveal several
important features. Firstly, the \nii\ emission is characteristic of
an expanding, hollow shell of gas whereas the \ha\ emission appears to
come from a large fraction of the nebular volume which also encloses an inner
\heii\ emitting  volume. This is consistent with previous \oiii\
observations of a double shell structure. This is not unexpected in
terms of ionisation stratification, but does pose serious problems if
the PN evolution is interpreted solely in terms of the interacting stellar
winds models; where the very hot and tenuous wind from the central
star is predicted to evacuate a hot bubble whose gas pressure 
drives and supports
the much more dense but relatively slowly expanding visible PN shell.

Incidentally, with an \nii\ expansion velocity of 35 \kms\ 
and by adopting the Benedict
\etal (2003) parallax distance of $420\pm 60$ pc and an angular
radius along the minor axis of $3\arcmin$ we get a true radius of 0.36
pc and a dynamical age for this inner bright structure of about 9000
years which must then be a lower limit for the true age of the \nii\ shell.

The Dumbbell could though, as a well-evolved PN, have a star no longer
producing the fast wind when in the white dwarf cooling region. 
In fact, Cerruti--Sola \& Perinotto (1985) and   
Patriarchi \& Perinotto (1991) fail to detect
any fast wind from the central star in the IUE observations.
In this
case the two--winds model may have been valid in the early stages 
of the Dumbbell's evolution as the star crossed the HR diagram 
horizontally. With the later decline of the wind the bubble pressure
would decrease and no longer balance the pressure of the expanding shell
of ionized gas with a consequent acceleration of its inner layers
towards the nebular centre. The increase in expansion velocities
from the core to the \nii\ emitting shell reported here are
consistent with this viewpoint.

It is also desirable to consider the alternative
possibility which involves `mass--loading' of the fast wind as it percolates
through the interior clumpy medium (Hartquist \etal
1986). Evidence for the presence of neutral globules (as also
seen in the Helix PN -- Meaburn \etal 1998) is shown in Meaburn
\& L{\'o}pez (1993) and in the expanding CO emitting shell
observed by Huggins \etal (1996).  It could be this slower
mass--loaded momentum--conserving outflow that is seen as  the \nii\ 
emitting outer shell,
itself clumpy. The `V--shaped' flows arrowed in Fig. 3 could
themselves be generated as this wind encounters further neutral
globules (see Steffen and  L{\'o}pez 2004 for numerical
simulations of this process). 

The possibility that the shape of the halo (see Fig. 1)
has been modified by the passage of the nebula through the
ambient interstellar medium (as suggested by Papamastorakis et al 1993)
has been reinforced by the recent proper motion measurements of Bendict et al 
(2003).These predict a tangential velocity of 37~\kms\ along 
PA 21 $\pm$ 1\deg\ (Fig. 1a)
of the central star which is in exactly
the direction expected if this motion has influenced the shaping
of the halo (see Fig. 1a). As differential 
galactic rotation only gives a tangential 
velocity
component of -8~\kms\ and radial velocity component of 5~\kms (to be compared 
with \vsys\ = -41 \kms -- see Fig. 4)
at this galactic longitude (l = 60.8\deg) the nebula must then
have a substantial velocity with respect to its  ambient medium.
The low turbulent velocities ($\leq$~20 \kms) 
reported here on the inner edge of the NE part of 
the clumpy halo is consistent with its origin in the AGB superwind.

\section{Conclusions}

The bright \nii\ emitting shell of the Dumbbell nebual 
is shown to be expanding radially
at 35 \kms\ along its minor axis but to be `quasi--ellipsoidal' or 
`barrel--shaped' along
its major axis.

The \ha\ emission from this bright  region is more diffuse and seems
to fill much of the  volume inside, and including, the outer \nii\ shell.
This is confirmed by the presence of an inner \oiii\ shell 
(expanding at 12 \kms)
surrounding an
\heii\
emitting  volume expanding at $\leq$~7~\kms.

The detailed kinematics of the \nii\ shell are characterised by 
`V-shaped' outflows in the pv arrays which are around 10\arcsec\ across.
It proposed that these could be formed as the expanding \nii\ emitting
shell (possibily composed of a mass--loaded wind) encounters slow moving
dense globules.

Line widths of $\approx$~20~\kms\ from the knotty, 
faint outer halo is observed 
just outside the bright nebulosity. An expansion velocity for this
outer halo of $\leq$~10 \kms\ is indicated in which case the bright
\nii\ emitting 
shell is colliding with a differential velocity of $\geq$~25 \kms.

The shape of the halo appears to have been affected by the
passage of the nebula through the ambient interstellar medium.

The present data indicate that the global kinematic pattern
of the Dumbbell nebula cannot be understood in simple terms
from the expected behaviour predicted by the interacting winds model.
Since the stellar wind from the central star seems to have ceased or
declined substantially, lack of inner support for bubble pressure could
have
modified the shell expansion pattern, producing fall back acceleration,
in agreement with the observations.

\acknowledgements

JM wishes to thank the staff of the WHT (La Palma) for their help
during these  observations.
This research has made use of data (POSS--II) obtained through 
the STScI Center
Online Service (CASB), provided by the California Institute of Technology.
Skinakas Observatory is a collaborative project of the University of
Crete, the Foundation for Research and Technology-Hellas and
the Max-Planck-Institut f\"ur Extraterrestrische Physik.JAL is grateful
to the Royal Society and Academia Mexicana de Ciencias 
for financing his July 2004 stay in Manchester.


\begin{thebibliography}{}

\bibitem{}
Balick, B. 1987, AJ, 94, 671

\bibitem{}
Balick, B., Gonzalez, G., Frank, A. \& Jacoby, G. 1992, ApJ, 392, 582

\bibitem{}
Balick, B. \& Frank, A. 2002, Annu. Review Astron. \& Astrophys., 40, 439.

\bibitem{}
Benedict, G. F. et al. 2003, AJ, 126,2549.

\bibitem{}
Boumis, P., Mavromatakis, F., Paleologou, E.V. \& Becker, W. 2002, A\&A,
396, 225.

\bibitem{}
Cerruti--Sola, M. \& Perinotto, M. 1985, ApJ, 291, 237.

\bibitem{}
Chu, Y.-H, Jacoby, G. H. \& Arendt, R. 1987, ApJS, 64, 529

\bibitem{}
Chu, Y.-H., Kwitter, K. B. \& Kaler, J. B. 1993, AJ, 106, 650

\bibitem{}
Frank, A.
1999, New Astr. Rev. 43 31

\bibitem{}
Goudis, C., McMullan, D., Meaburn, J., Tebbutt, N. J. \& Terrett, D. L. 1978, 
MNRAS, 182, 13.

\bibitem{}
Harris, H. C., Dahn, C. C., Monet, D. G. \&
Pier, J. R. 1997, in IAU Symp. 180, Planetary Nebulae, ed. H. J. Habing \&
H. J. G. L. M. Lamers (Dordrecht: Kluwer), 40.

\bibitem{}
Hartquist, T., Dyson, J. E., Pettini, M. \& Smith, L. 1986, MNRAS, 221, 715.

\bibitem{}
Huggins, P. J., Bachiller, R., Cox, P. \& Forveille, T. 1996, A\&A, 315, 284

\bibitem{}
Kahn, F. \& West, K., 1985, MNRAS 212, 837.

\bibitem{}
Kastner, J. H., Weintraub, D. A., Gatley, I., Merrill, K. M. \&
Probst, R. G. 1996, ApJ, 462, 777

\bibitem{}
Kwok, S., Purton, C. R. \& Fitzgerald, P. M., 1978, ApJ, 219, L125.

\bibitem{}
Lasker, B. M., Russel, J. N., \& Jenkner, H., 1999, in the HST Guide Star
Catalog, version 1.1-ACT, The Association of Universities for Research in
Astronomy, Inc.

\bibitem{}
Maury, A. \& Acker, A. 1990, Workshop Planetary Nebulae, A. Acker ed.,   
Strasbourg - Col de Steige

\bibitem{}
Manchado, A., Guerrero, M. A., Stanghellini, L. \& Serra--Ricart, M.
1996, in the IAC Morphological Catalogue of the Northern Galactic
Planetary Nebulae (IAC)

\bibitem{}
Meaburn, J., Blundell, B., Carling, R., Gregory, D. E., Keir, D. F. \&
Wynne C. G. 1984, MNRAS, 210, 463

\bibitem{}
Meaburn, J., Christopoulou, P.-E. \& Goudis C. D. 1992, MNRAS, 256, 97

\bibitem{}
Meaburn, J. \& L{\'o}pez, J. A. 1993, MNRAS, 263, 890

\bibitem{}
Meaburn, J., Clayton, C. A., Bryce, M., Walsh, J. R., Holloway,
A. J. \& Steffen, W. 1998, MNRAS, 294, 201

\bibitem{}
Millikan, A. G. 1974, AJ, 79, 1259

\bibitem{}
Miranda, L. F. \& Solf, J. 1992, A\&A, 260, 397

\bibitem{}
Moreno-Corral, M. A., L{\'o}pez-Molina, M. G. \& Vazquez, R. G. A. 1992,
RMAA, 24, 151.

\bibitem{}
Napiwotzki, R., 1999, A\&A, 350, 101.

\bibitem{}
O'Dell, C. R., 1998, AJ, 116, 1346.

\bibitem{}
O'Dell, C. R., McCullough, P. R. \& Meixner, M., 2004, AJ, 128, 2339.

\bibitem{}
Papamastorakis, J., Xilouris, K. M. \& Paleologou, E. V. 1993, A\&A, 279, 536.

\bibitem{}
Patriarchi, P. \& Perinotto, M. 1991, A\&A Suppl. Ser., 91, 325. 

\bibitem{}
Pier, J. R., Harris, H. C., Dahn, C.C. \& Monet, D. G. 1993, in Weinberger R., 
Acker A. A., eds, Proc. IAU Symp. 155, Planetary Nebulae. Kluwer, 
Dordrecht, 175



\bibitem{}
Sahai, R. \& Trauger, J.T., 1998, AJ, 116, 1357.

\bibitem{}Steffen W. \& L{\'o}pez, J. A. 2004, ApJ,612, 319.

\bibitem{}
Zuckerman, B. \& Gatley, I. 1988, ApJ, 324, 501

\end{thebibliography}
\end{document}